\documentclass[sigconf]{acmart}
\usepackage[ruled,linesnumbered,vlined]{algorithm2e}
\usepackage{multirow}
\usepackage{subfigure}
\usepackage{colortbl}
\usepackage{enumitem}

\AtBeginDocument{%
  }

\copyrightyear{2024}
\acmYear{2024}
\setcopyright{rightsretained} 
\acmConference[CIKM '24] {Proceedings of the 33rd ACM International Conference on Information and Knowledge Management}{October 21--25, 2024}{Boise, ID, USA.}
\acmBooktitle{Proceedings of the 33rd ACM International Conference on Information and Knowledge Management (CIKM '24), October 21--25, 2024, Boise, ID, USA}
\acmISBN{979-8-4007-0436-9/24/10}
\acmDOI{10.1145/3627673.3679537}

\usepackage{etoolbox}
\makeatletter
\patchcmd{\maketitle}{\@copyrightpermission}{
   \begin{minipage}{0.3\columnwidth}
     \href{https://creativecommons.org/licenses/by/4.0/}{\includegraphics[width=0.90\textwidth]{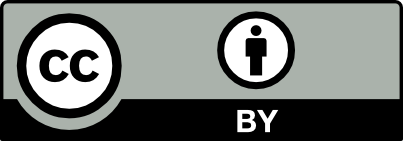}}
   \end{minipage}\hfill
   \begin{minipage}{0.7\columnwidth}
     \href{https://creativecommons.org/licenses/by/4.0/}{This work is licensed under a Creative Commons Attribution International 4.0 License.}
   \end{minipage}
   \vspace{5pt}
}{}{}

\makeatother

\begin{document}

% \author{Anonymous Author(s)}

\author{Zhiwei Yang}
\affiliation{%
   \institution{Institute of Information Engineering, Chinese Academy of Sciences, \& School of Cyber Security, University of Chinese Academy of Sciences,}
   \country{Beijing, China}}
\email{yzw377986686@gmail.com}

\author{Yuecen Wei}
\affiliation{%
   \institution{Beihang University}
   \country{Beijing, China}}
\email{weiyc@buaa.edu.cn}

\author{Haoran Li}
\affiliation{%
   \institution{Hong Kong University of Science and Technology,}
   \country{Hongkong, China}}
\email{hlibt@connect.ust.hk}

\author{Qian Li}
\affiliation{%
   \institution{Beijing University of Posts and Telecommunications,}
   \country{Beijing, China}}
\email{li.qian@bupt.edu.cn}

\author{Lei Jiang}
\affiliation{%
   \institution{Institute of Information Engineering, Chinese Academy of Sciences,}
   \country{Beijing, China}}
\email{jianglei@iie.ac.cn}
\authornote{Corresponding author}

\author{Li Sun}
\affiliation{%
   \institution{North China Electric Power University}
   \country{Beijing, China}}
\email{ccesunli@ncepu.edu.cn}

\author{Xiaoyan Yu}
\affiliation{%
   \institution{Beijing Institute of Technology}
   \country{Beijing, China}}
\email{xiaoyan.yu@bit.edu.cn}

\author{Chunming Hu}
\affiliation{%
   \institution{Beihang University}
   \country{Beijing, China}}
\email{hucm@buaa.edu.cn}

\author{Hao Peng}
\affiliation{%
   \institution{Beihang University,}
   \country{Beijing, China}}
\email{penghao@buaa.edu.cn}

\renewcommand{\shortauthors}{Zhiwei Yang, et al.}

\theoremstyle{definition}
\newtheorem{define}{Definition}[]

\title{Adaptive Differentially Private Structural Entropy Minimization for Unsupervised Social Event Detection}

\begin{abstract} 
Social event detection refers to extracting relevant message clusters from social media data streams to represent specific events in the real world.
Social event detection is important in numerous areas, such as opinion analysis, social safety, and decision-making.
Most current methods are supervised and require access to large amounts of data. These methods need prior knowledge of the events and carry a high risk of leaking sensitive information in the messages, making them less applicable in open-world settings.
Therefore, conducting unsupervised detection while fully utilizing the rich information in the messages and protecting data privacy remains a significant challenge.
To this end, we propose a novel social event detection framework, ADP-SEMEvent, an unsupervised social event detection method that prioritizes privacy. 
Specifically, ADP-SEMEvent is divided into two stages, i.e., the construction stage of the private message graph and the clustering stage of the private message graph. 
In the first stage, an adaptive differential privacy approach is used to construct a private message graph. 
In this process, our method can adaptively apply differential privacy based on the events occurring each day in an open environment to maximize the use of the privacy budget. 
In the second stage, to address the reduction in data utility caused by noise, a novel 2-dimensional  structural entropy minimization algorithm based on optimal subgraphs is used to detect events in the message graph. 
The highlight of this process is unsupervised and does not compromise differential privacy. 
Extensive experiments on two public datasets demonstrate that ADP-SEMEvent can achieve detection performance comparable to state-of-the-art methods while maintaining reasonable privacy budget parameters.
\end{abstract}



\maketitle
\section{Introduction}

\begin{figure}[t]
    \centering
    \includegraphics[width=0.445\textwidth]{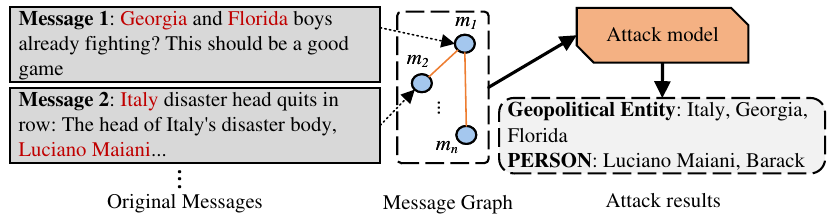}
    \vspace{-3mm}
    \caption{A toy example of an attack demonstrating the potential leakage of sensitive information such as persons or geopolitical entities.}
    \vspace{-5mm}
    \label{toy example}
    \Description{toy example.}
\end{figure}

Social event detection refers to extracting relevant clusters of messages from social media message corpora or social message streams to represent specific events in the real world.
Privacy leakage is a crucial concern today and also exists in social event detection tasks. 
Social events occur daily, often containing rich information regarding collective social behaviors that widespread public attention. 
Therefore, the task of social event detection holds significant implications for sentiment analysis ~\cite{beck2021investigating}, product recommendations, and decision-making ~\cite{liu2017event}, distinguishing between real and fake news ~\cite{mehta2022tackling}, as well as monitoring and managing social crises ~\cite{pohl2015social}. 
% For these reasons, social event detection has become a focal point in social media mining over the past decade, garnering increasing attention from academia and industry. 

% Introduction to traditional methods and their limitations
Most early methods for social event detection ~\cite{zhao2011comparing,amiri2016short,feng2015streamcube,morabia2019sedtwik,gao2016collaborative,liu2017cpmf,wang2017neural,yan2015probabilistic} primarily relied on extracting attributes from text content and subsequently applying traditional statistical or machine learning methods for detection. 
However, these detection methods exhibited low accuracy and were not widely adopted due to limitations such as insufficient richness in extracted text features and inadequate interaction between different attributes.
% Introduction to GNN methods and their limitations
Later, with the development of Graph Neural Networks (GNNs), GNN-based methods for social event detection gradually gained popularity. 
GNNs can effectively facilitate direct interactions between nodes and neighbors in a graph, leading many researchers to improve performance by leveraging GNN models~\cite{peng2019fine,peng2021streaming,cao2021knowledge,peng2022reinforced}. 
However, most methods have been tested under the supervised learning assumption that social event categories are predefined.
The supervised assumption conflicts with the acquisition of social events in the real world, i.e., the total number of event categories in the real world frequently varies, making it impossible to predefine the total number of social event categories.
Therefore, supervised GNN-based methods may not perform well in open environments.
% Introducing the latest unsupervised approach, noting that it fails to protect privacy
Recently, QSGNN ~\cite{ren2022known} and HISEvent ~\cite{cao2024hierarchical} proposed to enhance social event detection in open environments. 
QSGNN successfully implemented social event detection in open environments using a pseudo-labeling method.
HISEvent achieved state-of-the-art (SOTA) unsupervised social event detection using Structural Entropy ~\cite{li2016structural} (SE) combined with hierarchical clustering. 
Both of them can be applied effectively in the real world, but they still suffer from a serious problem, which is the potential leakage of sensitive private data due to their lack of protection for sensitive attributes within the data. 
Graph-based social event detection methods typically use various additional information to enrich the graph's connectivity ~\cite{peng2019fine,ren2022known,cao2024hierarchical}. 
Though enriching edges can improve detection performance, it can also lead to unintentional privacy leakage of user privacy data through member inference attacks, attribute inference attacks, and other methods, Figure ~\ref{toy example} shows a toy example of the attack. 
To the best of our knowledge, among all the methods mentioned above, there is no social event detection model that can handle both unsupervised learning and privacy protection.  
We believe that both aspects are necessary.

% Introduce our work and list contributions
In this work, we propose a novel framework, \textit{\textbf{A}daptive \textbf{D}ifferentially \textbf{P}rivate \textbf{S}tructural \textbf{E}ntropy \textbf{M}inimization for Unsupervised Social \textbf{Event} Detection} (ADP-SEMEvent), which simultaneously achieves outstanding social event detection performance with privacy protection on unsupervised learning. 
\textbf{To respect privacy}, inspired by research on differential privacy ~\cite{dwork2006differential} (DP), we implement an adaptive differentially private strategy in ADP-SEMEvent.
We use a mixed sensitivity strategy to achieve adaptive DP, enabling our model to automatically select the most noise level based on different message sets on different days. 
This adaptive nature effectively balances privacy protection and detection accuracy. 
Specifically, we calculate the similarity between nodes using the adaptive differentially private strategy. Then we explore the edges connectivity of the graph using 1-dimensional Structural Entropy (1D SE) and relevant attributes, ultimately obtaining a private message graph.
\textbf{For unsupervised learning}, we employ SE from information theory as a crucial criterion during the clustering process, an indicator for evaluating the amount of information contained within a graph. 
Subsequently, to overcome the current 2-dimensional Structural Entropy (2D SE) minimization algorithm's performance and efficiency limitations, we designed a 2D SE minimization method based on optimal subgraphs to cluster the message graph for social event detection.
The source code is available on GitHub\footnote{https://github.com/SELGroup/ADP-SEMEvent}.

Our experiments on two public Twitter datasets demonstrate that ADP-SEMEvent outperforms baselines for both open and closed set settings and remains competitive with state-of-the-art approaches within a certain range of privacy budget settings. 
To the best of our knowledge, we are the first study to implement unsupervised social event detection with differentially private guarantees. 
Our contributions are as follows: 
1) We incorporate an adaptive differentially private strategy into the process of social event detection, striking a balance between privacy protection and detection accuracy. 
2) We propose a novel 2D SE minimization algorithm based on optimal subgraphs to partition the graph and use it for clustering the message graph to conduct unsupervised social event detection. 
3) Extensive experiments conducted on two public Twitter datasets demonstrate that ADP-SEMEvent, under a reasonable privacy budget, can achieve performance comparable to current state-of-the-art models while preserving privacy.
\section{Preliminary}
\begin{figure*}[!htb]
    \centering
    \includegraphics[width=1\textwidth]{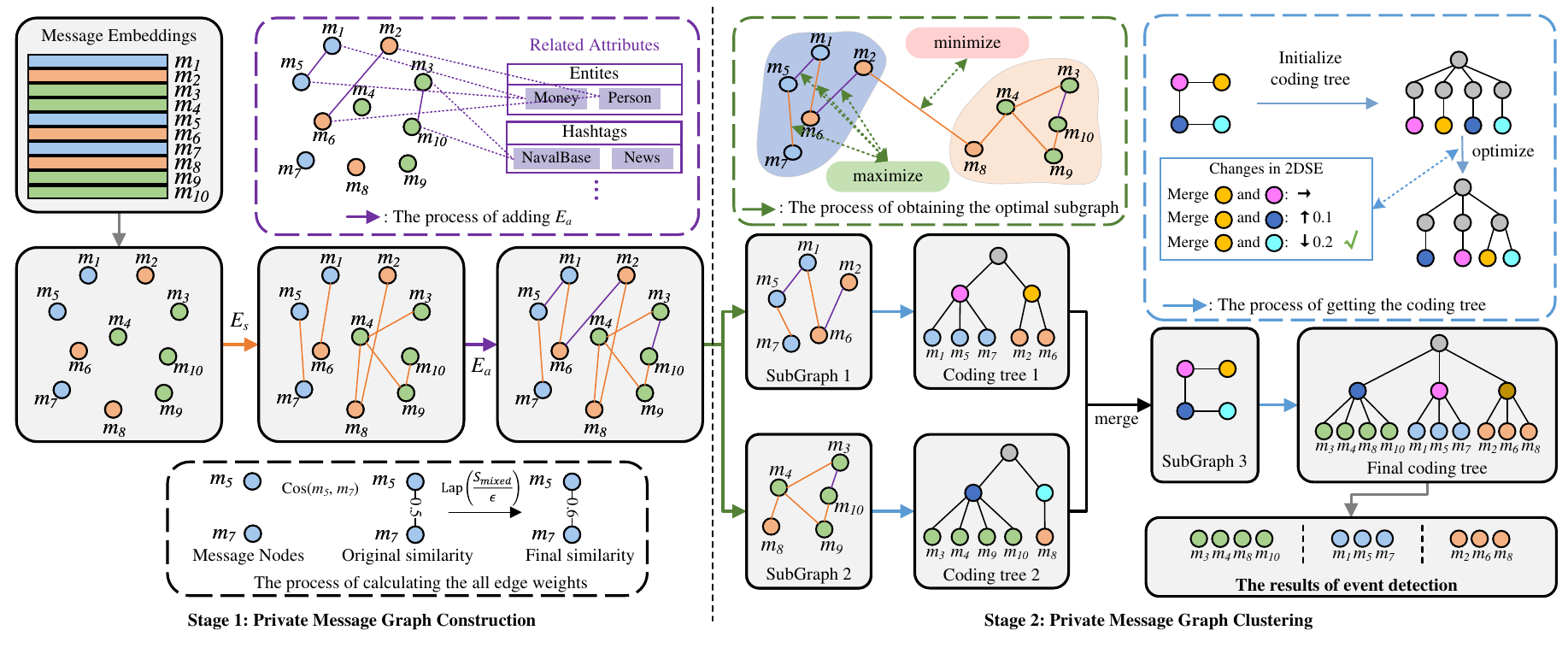}
    \vspace{-5mm}
    \caption{The proposed ADP-SEMEvent framework. ADP-SEMEvent consists of two stages: the private message graph construction stage (stage 1) and the private message graph clustering stage (stage 2). Messages with the same color represent the same cluster according to the ground truth labels; orange edges $E_s$ are derived from 1-dimensional structural entropy, and purple edges $E_a$ are derived from relevant attributes; arrows of specific colors indicate specific operations.}
    \label{framework}
    \vspace{-2mm}
\end{figure*}

\subsection{Differentially Private Message Graph}\label{Release}
The release of messages graph refers to the process of transforming a series of social messages set $ D = \{ m_1, \ldots, m_N \} $ (generally, using Pre-trained Language Models (PLMs) to convert the message into embeddings) into a message graph $ G = (V, E) $ using an algorithm $ R $. 
$ G $ is utilized for social event detection. 
Under the background of protecting privacy, we present the following definitions:

\begin{definition}\label{Differentially}(Differential Privacy ~\cite{dwork2006differential}). 
For algorithm $ R $, given social message sets $ D $ and $ D' $ ($ D $ and $ D' $ are neighboring datasets that differ by only one message), and any output $\mathcal{G} \subseteq \text{range}(D)$. 
For any $ \epsilon > 0 $, if the algorithm $ R $ satisfies:
\begin{equation}\label{DP_eq}
\Pr[R(D) \subseteq \mathcal{G}] \leq e^{\epsilon} \Pr[R(D') \subseteq \mathcal{G}] + \delta, 
\end{equation}
we say that the algorithm $ R $ satisfies $ (\epsilon, \delta) $-DP, where $ \epsilon $ is the privacy budget. 
Particularly, if $ \delta = 0 $, the algorithm $ R $ satisfies $ \epsilon $-DP. 
$ R $ can protect sensitive information contained in the messages during the release of the message graph.
\end{definition}

\begin{definition}\label{Sensitivity}(Sensitivity).
Consider a function $R$ whose input is a message set and whose output is in $ \mathbb{R}^k $. 
The sensitivity $ S_R $ of function $ R $ is defined as:
\begin{equation}  \label{S_R}
S_R = \max \| R(D) - R(D') \|_1. 
\end{equation} 
Sensitivity is used to determine the magnitude of random perturbation noise to protect private information, and the sensitivity includes global sensitivity ~\cite{dwork2006calibrating} ($S_{global}$) and local sensitivity ~\cite{nissim2007smooth} ($S_{local}$).
\end{definition}

\begin{definition}\label{Laplace}(Laplace Mechanism for Message Similarity). 
We measure the similarity between any two messages using the cosine similarity, then
% \begin{equation}
% \text{Cos}(m_i, m_j) = \frac{m_i \cdot m_j}{|m_i| \cdot |m_j|}. 
% \end{equation}
% the Laplace mechanism ~\cite{dwork2006calibrating} is applied to perturb the similarity with noise:
the Laplace mechanism ~\cite{dwork2006calibrating} is applied to generate the noise and perturb the similarity as:
\begin{equation} \label{cos_lap}
\text{Cos}_{Lap}(m_i, m_j) = \text{Cos}(m_i, m_j) + \text{Lap}(\dfrac{S}{\epsilon}), 
\end{equation}
where $S$ is typically calculated using global sensitivity, but we designed a novel mixed sensitivity approach for $S$.
(described in Section ~\ref{Sec_1dse}). 
$\text{Lap}({S}/{\epsilon})$ denotes sampling from a Laplace distribution with mean 0 and scale factor $S$.
\end{definition}

\subsection{Structural Entropy}\label{SE}
Social event detection can be conducted by constructing classification or clustering models on the messages graph. 
Consistent with Cao et al.'s ~\cite{cao2024hierarchical} approach, we achieve unsupervised graph clustering using SE minimization.
We provide the relevant definitions below:
\begin{definition}\label{codetree}(Encoding Tree of Messages Graph ~\cite{li2016structural, li2015discovering}).
The coding tree $T$ of the messages graph $ G = (V, E) $ is a hierarchical clustering partition of $ G $ (illustrated as the blue rounded dashed line box in Figure ~\ref{framework}).
The coding tree $T$ includes all message nodes as leaf nodes. Each node $\alpha$ in $T$ corresponds to a partitioning of message nodes, with the set $ T_{\alpha} = {v_{a}^1, \ldots, v_{a}^j} $, representing the successor nodes of $ \alpha $. The root node $ \lambda $ of $T$ has the set $ T_{\lambda} = V $, indicating no partitioning.
For each node $\alpha$ in $T$ (excluding $\lambda$), the height $h(\alpha)$ is one less than that of its parent node. The root node $ \lambda $ has a height of 0. The height of $T$ is the maximum height among all nodes in $T$.
\end{definition}
% \begin{definition}\label{codetree}(Encoding Tree of Messages Graph ~\cite{li2016structural, li2015discovering}).
% The coding tree $T$ of the messages graph $ G = (V, E) $ is a hierarchical clustering partition of $ G $ (the blue rounded dashed line box in Step 2 of Figure ~\ref{framework} is an example of a coding tree). 
% It satisfies the following properties:
% 1) All message nodes are included in the coding tree $T$ and are leaf nodes. 
% 2) Each node $\alpha$ in $T$ corresponds to a partitioning of the message nodes, with the partitioning set $ T_{\alpha} = \{v_{a}^1, \ldots, v_{a}^j\} $, where $ v_{a}^i $ represents the message nodes in the successor nodes of $ \alpha $. Specifically, for the root node $ \lambda $ of $T$, $ T_{\lambda} = V $, indicating no partitioning. 
% 3) For each node $ \alpha $ in $T$ (other than $ \lambda $), the height is denoted as $ h(\alpha) $, with the height of its parent node $\alpha^-$ as $ h(\alpha^-) = h(\alpha) - 1 $. In particular, \( h(\lambda) = 0 \). 
% The height of $T$, denoted as $ h(T) $, is the maximum height across all nodes in $ T $: $ h(T) = \underset{\alpha \in {T}}{\max}  \{h(\alpha)\} $. 
% \end{definition}

\begin{definition}\label{1-dimensional}(1D SE).
For a messages graph $G = (V, E)$ with $n$ vertices, its 1D SE is defined as:
\begin{equation}\label{1dse}
{H}^{(1)}(G) = - \sum_{i=1}^{n} \dfrac{d_i}{2m}\text{log}\dfrac{d_i}{2m}, 
\end{equation}
where $d_i$ represents the weighted degree of node $i$, and $m$ denotes the sum of the weighted degrees of all nodes.

\end{definition}

\begin{definition}\label{2-dimensional}(2D SE).
For a messages graph $G = (V, E)$ with $n$ vertices, $P = \{p_1, \ldots, p_L\}$ is a partition of $V$, $p_j$ represents an event cluster in the message graph. The 2D SE is defined as:
\begin{equation}
{H}^{(2)}(G) = - \sum_{j=1}^{L} \dfrac{V_j}{m} \sum_{i=1}^{n_j} \dfrac{d_i^{j}}{V_j} \text{log}_2 {\dfrac{d_i^{j}}{V_j}} - \sum_{j=1}^{L}\dfrac{P_{cut_j}}{m}\text{log}_2 \dfrac{V_j}{m},  
\end{equation}  
where $n_j$ is the number of nodes in partition $p_j$, $d_i^j$ is the weighted degree of the $i$th node in $p_j$, $V_j$ is the sum of the weighted degrees of all nodes in partition $p_j$, and $P_{\text{cut}_j}$ is the sum of the weights of the cut edges in $p_j$.
\end{definition}

\section{Methodology}
\begin{algorithm}[!t]
    \caption{Message Graph Synthesis under Adaptive Differentially Private Strategy}
    \label{Algorithm 1}
    \leftskip=0pt %
    \setcounter{AlgoLine}{0} % 重置行号计数器
    \KwIn{social message embedding set $M = \{m_1, m_2, m_3, ..., m_N\}$; maximum number of neighbors $k_{max}$.}
    \KwOut{a message graph $G = (V, E) $.}
    \BlankLine
    Set $V = \{1, 2, 3, ... ,N\}$; \\
    Get similarity $\leftarrow$ Eq.~\eqref{Cos_lap}; \\
    Set $k = 1$, $E_s = \emptyset$, $E_s^{'} = \emptyset$, $\text{SE} = +\infty$; \\
    \While{True }{
        initialize: $G = (V, E_s) $, $E_s^{'} = \emptyset$;\\
        \For{each $v_i \in V$}{
            $E_s^{'} \overset{\text{add to}}{\leftarrow}  \text{Let $v_i$ connect the $k$ largest similarity nodes}$;
        }
        ${H}^{(1)}(G)$ $\leftarrow$ Eq.~\eqref{1dse};\\
        \eIf{${H}^{(1)}(G) < \text{SE}$ \textbf{and} $k < k_{max}$}{
            $\text{SE} = {H}^{(1)}(G)$;\\
            $E_s = E_s^{'}$;\\
            $k++$;\\
        }{
            \textbf{break};
        }
    }
    Set $E_a = \emptyset$;\\
    \For{each $i \in [1, \text{size}(V)]$}{
        \For{each $j \in [i+1, \text{size}(V)]$}{
            \If{$v_i$ and $v_j$ have shared attributes}{
                $E_a \overset{\text{add to}}{\leftarrow} \text{similarity($v_i$, $v_j$)}$;
            }
        }
    }
    $E = E_s + E_a$;\\
    $G = (V, E)$;\\
    \Return $G$.
\end{algorithm}

In this section, we provide a detailed description of ADP-SEMEvent. 
The entire framework is illustrated in Figure ~\ref{framework}.
The private message graph construction stage primarily utilizes our proposed adaptive differentially private strategy (introduced in Section ~\ref{Sec_Dynamic}), combined with 1D SE and relevant attributes for constructing the private message graph (discussed in Section ~\ref{Sec_1dse}). 
The private message graph clustering stage involves a 2D SE minimization algorithm based on optimal subgraphs (explained in Section ~\ref{Sec_2dse}).

\subsection{Problem Formalization}\label{Sec_Problem}
Given a series of social messages $m_1, \ldots , m_N$ forming a message graph $G = (V, E) $, where the node set $V = \{m_1, \ldots , m_N \}$ and the edge set $E$ represents the relevance between messages. 
The task of social event detection requires partitioning the message graph $G$ into several partitions, denoted as $\{e_1, \ldots , e_M \}$, where $e_i \subset V$, $ e_i \cap e_j = \emptyset $,  $ e_1 \cup \ldots \cup e_M = V $. 
The partitions represent $M$ detected message clusters. 
Specifically, the message graph $G$ contains private information and poses a risk of being attacked.

\subsection{Adaptive Differentially Private Strategy}\label{Sec_Dynamic}
This section will introduce our mixed sensitivity differentially private strategy and provide proof that this strategy satisfies $(\epsilon, \delta)$-DP in the worst-case scenario.

\subsubsection{Mixed Sensitivity Strategy}\label{Mixed_S}
To prevent the waste of privacy budget $\epsilon$ caused by the global sensitivity being much greater than the local sensitivity in a specific data set, we design a mixed sensitivity strategy to prevent the waste of privacy budget and the addition of unnecessary noise.

In Equation ~\ref{S_R}, $S_{global}$ refers to the maximum difference between the results of a query for any two different data records in any data set. 
$S_{global}$ is independent of the data set and is related only to the calculation method itself. 
$S_{local}$ removes the restriction of any data set from the definition of global sensitivity and only considers the current data set. 
$S_{local}$ can be obtained through a limited number of calculations. 
In general, we can not directly use $S_{local}$ in the Laplace mechanism because if a malicious attacker knows the local sensitivity of a function $ f $ on a specific dataset, they can still infer some information about the dataset ~\cite{nissim2007smooth}. 
Therefore, $S_{local}$ must be smoothed to prevent the malicious attacker from deducing the local sensitivity. 
This is achieved using smooth sensitivity ~\cite{nissim2007smooth}, denoted as $ S_{smooth} $), which is calculated as:
\begin{equation}
S_{smooth} = 2\text{exp}{(-\dfrac{\epsilon}{2} \cdot \log(\dfrac{2}{\delta}))} \cdot S_{local}, 
\end{equation}
where $ \delta $ is typically set to ${1}/{|D|^2}$($|D|$ is the size of $D$). 

Then, we design the mixed sensitivity ($S_{mixed}$) strategy. 
% The core idea of this strategy is to select the smaller value between smooth sensitivity and global sensitivity for the Laplace mechanism, ensuring both data security and model performance. 
$S_{mixed}$ is calculated as follows:
\begin{equation}\label{M-MDP}
S_{mixed} = \min \{ S_{global}, S_{smooth} \}. 
\end{equation}
With the privacy budget $\epsilon$ unchanged, our strategy can automatically determine fitness sensitivity based on daily events to achieve adaptive noise perturbation.

\subsubsection{Differential Privacy Proof}\label{Proof_DP}
We prove that our strategy worst-case satisfies $(\epsilon,\delta)$-DP. 
This entails demonstrating that global sensitivity satisfies $\epsilon$-DP and that smooth sensitivity satisfies $(\epsilon,\delta)$-DP.

For $S_{smooth}$, Nissim et al. ~\cite{nissim2007smooth} have provided a rigorous proof that using $S_{smooth}$ for the Laplace mechanism always satisfies $(\epsilon,\delta)$-DP, regardless of whether global sensitivity can be found. 
Where $\delta$ represents the probability of failure, and in our strategy, $\delta = {1}/{|D|^2}$, is a small probability. 
For $S_{global}$, we provide the following proof:
\vspace{-2.5mm}
\begin{proof}
For $D$ and $D'$, $P_D$ and $P_{D'}$ represent the probability distributions of message similarities calculated on their respective datasets. 
For any two messages $m_i$ and $m_j$ in the dataset, the similarity is represented as $ z = \text{Cos}(m_i, m_j) + Y $, where $Y$ follows Laplace distribution with scale $ {S_{mixed}}/{\epsilon} $ and mean 0. 
Therefore, we have:
\begin{equation}\label{P_D}
P_D(z) = \frac{\epsilon}{2S_{mixed}} \exp(-\frac{\epsilon|\text{Cos}_D(m_i, m_j) - z|}{S_{mixed}}) 
\end{equation}
\begin{equation}\label{P_D2}
P_{D'}(z) = \frac{\epsilon}{2S_{mixed}} \exp(-\frac{\epsilon|\text{Cos}_{D'}(m_i, m_j) - z|}{S_{mixed}}). 
\end{equation}
Next, verify whether it complies with differential privacy through distribution:
\begin{align}\label{proof}
\frac{P_D(z)}{P_{D'}(z)} & = \frac{\exp\left(-\frac{\epsilon|\text{Cos}_D(m_i, m_j) - z|}{S_{mixed}}\right)}{\exp\left(-\frac{\epsilon|\text{Cos}_{D'}(m_i, m_j) - z|}{S_{mixed}}\right)} \\
                        & = \exp\left(\frac{\epsilon(|\text{Cos}_{D'}(m_i, m_j) - z| - |\text{Cos}_D(m_i, m_j) - z|)}{S_{mixed}}\right)\\
                        & \leq \exp\left(\frac{\epsilon(|\text{Cos}_{D'}(m_i, m_j) - \text{Cos}_D(m_i, m_j)|)}{S_{mixed}}\right) \\
                        & \leq \exp\left(\epsilon\right).
\end{align}
\end{proof}
In the process of proving, the last two steps follow from the triangle inequality. 
According to Equation ~\ref{DP_eq}, our strategy satisfies $\epsilon$-DP. 
In summary, our strategy worst-case satisfies $(\epsilon,\delta)$-DP. 

\begin{algorithm}[!t]
    \caption{2D SE Minimization Based on Optimal Subgraphs}
    \label{Algorithm 2}
    \leftskip=0pt %
    \setcounter{AlgoLine}{0} % 重置行号计数器
    \LinesNumbered % 显示行号
    \KwIn{message graph $G = (V, E) $; Subgraph size $q$; initial partition $P = \{p_1, \dots ,p_m\}$.}
    \KwOut{Clustering results $C = \{c_{1}, \dots ,c_{n}\}$.}
    \BlankLine
    $C = P$;\\
    \While{True}{
        \tcp{Synthetic Super-graph based on partitions $C$}
        $V^\prime$ $\leftarrow$ The nodes in the same partition $c_{n}$ are combined into a super-node;\\
        $E^\prime$ $\leftarrow$ The edge weights between $V^\prime$ are summed by the original nodes;\\
        $G^\prime = (V^\prime, E^\prime)$ $\leftarrow$ $V^\prime$, $E^\prime$;\\
        \tcp{Get the optimal subgraphs}
        Calculate the number of subgraphs $k_{max}$ = $\left\lceil \frac{|C|}{q} \right\rceil$;\\
        \For{each $k$ $\in$ $ [1, k_{max}]$}{
            Set $g_{sub_k} = (V_{sub_k}, E_{sub_k})$, $V_{sub_k} = \emptyset$, $E_{sub_k} = \emptyset$;\\
            $g_{sub_k}$ $\overset{\text{add to}}{\leftarrow}$ The endpoint of the largest edge in $G^\prime$; \\
            \While{$|V_{sub_k}| < q$}{
                $g_{sub_k} \overset{\text{add to}}{\leftarrow}$ The node with the highest cut weight to current $g_{sub_k}$;
            }
            $G^\prime = G^{\prime} - g_{sub_k}$;
        }
        \tcp{2D SE minimization of optimal subgraphs}
        set $C = \emptyset$, $C_{last} = C$;\\
        \For{each $g_{sub_k}$ $\in$ $\{g_{sub_1}, \dots ,g_{sub_k}\}$}{
            $T$ $\leftarrow$ Construct an initial coding tree according to definition ~\ref{codetree};\\
            $T_{best}$ $\leftarrow$ Run vanilla 2D SE minimization to minimize 2D SE;\\
            $C \overset{\text{add to}}{\leftarrow}$ The partition corresponding to $T_{best}$;
        }
        \uIf{$C = C_{last}$}{
            $q = 2q$;
        }
        \uElseIf{$k_{max} = 1$}{
            \textbf{break};
        }
        \lElse{
            $C_{last} = C$
        }
    }
    \textbf{return} $C$.
\end{algorithm}

\subsection{Private Message Graph Synthesis}\label{Sec_1dse}
To ensure the generation of rich and private message graphs, we propose a method of message graph synthesis with edges $E_s$ based on SE and edges $E_a$ based on attribute. 
We employ our proposed adaptive differentially private strategy to protect privacy while computing edge weights.

We summarize our method into \textbf{Algorithm ~\ref{Algorithm 1}}, termed \textit{Message Graph Synthesis under Adaptive Differentially Private Strategy}. 
For $E_s$ (lines 3–14), we link each message to its $k$ nearest neighbors, where the distance utilizes cosine similarity under the adaptive differentially private strategy:
\begin{equation}\label{Cos_lap}
\text{Cos}_{Lap}(m_i, m_j) = \frac{m_i \cdot m_j}{|m_i| \cdot |m_j|} + \text{Lap}(\frac{S_{mixed}}{\epsilon}),
\end{equation}
where $m_k$ represents the semantic vector corresponding to the message node, obtained through PLMs (SBERT ~\cite{reimers2019sentence}). 
It is worth noting that, due to the cosine similarity's range being [-1, 1], $ S_{global} = 2 $. 
We gradually increase the number of neighbors from $k = 1$, then compute the 1D SE under different $k$ neighbors, selecting the linkage method with the smallest 1D SE as the final $E_s$. 
Specifically, to reduce algorithmic overhead, we greedily select the first occurrence of local minimum 1D SE as our target rather than using the globally optimal target. 
For $E_a$ (lines 15–19), we draw inspiration from the majority of GNN-based social event detection methods ~\cite{peng2019fine,cao2021knowledge,ren2022known,peng2022reinforced}, extracting common attributes of social events such as entities, user mentions, and user IDs. We set the edge weight between nodes $v_i$ and $v_j$, sharing the same attributes to the cosine similarity $ \text{Cos}_{Lap}(m_i, m_j) $ corresponding to the nodes. 
Finally, we combine $E_a$ and $E_s$ as all edges of the final private graph.

\subsection{Event Detection via 2D SE Minimization}\label{Sec_2dse}

\begin{figure}[!t]
    \centering
    \includegraphics[width=0.44\textwidth]{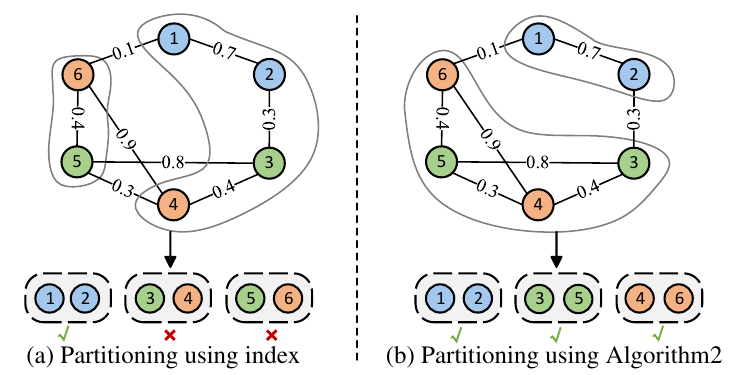}
    \vspace{-3mm}
    \caption{The impact of different ways of initializing partitions on the results.}
    \vspace{-4mm}
    \label{vs}
\end{figure}

To efficiently cluster on the noisy message graph $G$, we design an algorithm named \textit{2D SE minimization based on optimal subgraphs}. 
% This step can be seen as a clustering process for the nodes in the graph to group related messages into the same cluster and separate unrelated messages into different clusters. 
Existing approaches used a greedy approach (\textit{vanilla 2D SE minimization} ~\cite{li2016structural}) to repeatedly merge any two nodes in the encoding tree $T$ for 2D SE minimization, which is time-consuming on large graphs.
Later, an incremental method reduces runtime overhead by sequentially dividing large message graphs into multiple subgraphs~\cite{cao2024hierarchical}. 
Nevertheless, such methods addressed the problem of high time complexity, but simply dividing the subgraphs is unwise. 
The initial partitioning is crucial because it may force strongly related message nodes to be divided into different parts, ultimately preventing them from merging into the same cluster, as shown in Figure \ref{vs} (a). 
So, we prevent strongly correlated message nodes from being forcibly separated by constructing optimal subgraphs, as illustrated in Figure \ref{vs} (b).

Specifically, as shown in \textbf{Algorithm ~\ref{Algorithm 2}}, assuming a graph $G = (V, E)$, partition $P = \{p_1, \ldots, p_m\}$ (where $p_m \subset V, p_i \cap p_j = \emptyset$), and subgraph size $q$ as input. 
In each iteration, we first construct a super-graph based on the original graph and the initial partition (lines 3–5), treating nodes in the same partition as a single new node and using the sum of the cut edges between partitions as the edge weight between new nodes. 
Then, using a greedy approach, we obtain the optimal subgraph (lines 6–12), attempting to include edges with higher weights in the same subgraph as much as possible. 
Finally, we perform 2D SE minimization on each subgraph (lines 13–17), achieving the final clustering result based on optimizing each encoding tree.

\begin{table*}[!t]
    \vspace{-2mm}
    \caption{Open-set results on Event2012.}
    \vspace{-2mm}
    \centering
    \renewcommand\arraystretch{0.92}
    \resizebox{\linewidth}{!}{
    \begin{tabular}{c||c c|c c|c c|c c|c c|c c|c c}
        \toprule
        \multirow{2}{*}{Methods} &\multicolumn{2}{c|}{$M_1$} &\multicolumn{2}{c|}{$M_2$} &\multicolumn{2}{c|}{$M_3$} &\multicolumn{2}{c|}{$M_4$} &\multicolumn{2}{c|}{$M_5$} &\multicolumn{2}{c|}{$M_6$} &\multicolumn{2}{c}{$M_7$}\\
        \cline{2-15}
        &AMI &ARI &AMI &ARI &AMI &ARI &AMI &ARI &AMI &ARI &AMI &ARI &AMI &ARI \\
        \hline
        BERT &0.35 &0.03 &0.76 &0.65 &0.72 &0.45 &0.58 &0.19 &0.67 &0.36 &0.75 &0.45 &0.50 &0.07 \\ 
        SBERT&0.38 &0.03 &0.85 &0.73 &0.87 &0.68 &0.80 &0.36 &\textbf{0.85} &0.61 &0.83 &0.53 &0.61 &0.09 \\
        EventX &0.06 &0.01 &0.29 &0.45 &0.18 &0.09 &0.19 &0.07 &0.14 &0.04 &0.27 &0.14 &0.13 &0.02 \\
        KPGNN &0.37 &0.07 &0.87 &0.76 &0.74 &0.58 &0.64 &0.29 &0.71 &0.47 &0.79 &0.72 &0.51 &0.12\\
        QSGNN &0.41 &0.07 &0.80 &0.77 &0.76 &0.59 &0.68 &0.29 &0.73 &0.48 &0.80 &0.73 &0.54 &0.12\\
        HISEvent &0.44 &0.08 &\textbf{0.88} &0.79 &\textbf{0.94} &\textbf{0.95} &0.84 &0.50 &\textbf{0.85} &0.62 &\textbf{0.90} &0.86 &\textbf{0.68} &0.27 \\
        \hline
        DP$_G$-SEMEvent ($\epsilon$=10) &0.18 &0.05 &0.72 &0.72 &0.64 &0.69 &0.37 &0.21 &0.61 &0.58 &0.66 &0.67 &0.34 &0.20 \\
        DP$_G$-SEMEvent ($\epsilon$=15) &0.33 &0.07 &0.81 &0.84 &0.81 &0.85 &0.40 &0.50 &0.79 &0.78 &0.80 &0.80 &0.54 &0.23 \\
        ADP-Spectral Clustering ($\epsilon$=10)  &0.05 &0.07 &0.01 &0.02 &0.05 &0.08 &0.02 &0.10 &-0.01 &0.01 &0.02 &0.11 &0.12 &0.07\\
        ADP-Spectral Clustering ($\epsilon$=15) &0.08 &0.05 &0.02 &0.03 &0.10 &0.11 &0.02 &0.03 &0.05 &0.02 &0.03 &0.02 &0.13 &0.09 \\
        \hline
        ADP-SEMEvent ($\epsilon$=10) &0.22 &0.08 &0.77 &0.80 &0.73 &0.73 &0.51 &\textbf{0.63} &0.62 &0.59 &0.80 &0.83 &0.35 &0.25 \\
        ADP-SEMEvent ($\epsilon$=15) &0.35 &\textbf{0.09} &0.87 &\textbf{0.87} &0.89 &0.91 &0.73 &0.49 &\textbf{0.85} &\textbf{0.85} &0.89 &0.87 &0.57 &0.28 \\
        ADP-SEMEvent ($\epsilon$=None) &\textbf{0.46} &\textbf{0.09} &0.87 &\textbf{0.87} &0.92 &0.91 &\textbf{0.85} &0.55 &\textbf{0.85} &0.62 &\textbf{0.90} &\textbf{0.93}  &0.65 &\textbf{0.35}\\
        \toprule

        \toprule
        \multirow{2}{*}{Methods} &\multicolumn{2}{c|}{$M_8$} &\multicolumn{2}{c|}{$M_9$} &\multicolumn{2}{c|}{$M_{10}$} &\multicolumn{2}{c|}{$M_{11}$} &\multicolumn{2}{c|}{$M_{12}$} &\multicolumn{2}{c|}{$M_{13}$} &\multicolumn{2}{c}{$M_{14}$}\\
        \cline{2-15}
        &AMI &ARI &AMI &ARI &AMI &ARI &AMI &ARI &AMI &ARI &AMI &ARI &AMI &ARI \\
        \hline
        BERT &0.74 &0.51 &0.71 &0.34 &0.78 &0.55 &0.62 &0.26 &0.56 &0.31 &0.57 &0.13 &0.55 &0.24 \\
        SBERT &0.86 &0.65 &0.83 &0.47 &0.85 &0.62 &\textbf{0.82} &0.49 &0.85 &0.63 &0.70 &0.24 &0.77 &0.40  \\
        EventX &0.21 &0.09 &0.19 &0.07 &0.24 &0.13 &0.24 &0.16 &0.16 &0.07 &0.16 &0.04 &0.14 &0.10\\
        KPGNN &0.76 &0.60 &0.71 &0.46 &0.78 &0.70 &0.71 &0.49 &0.66 &0.48 &0.67 &0.29 &0.65 &0.42\\
        QSGNN &0.75 &0.59 &0.75 &0.47 &0.80 &0.71 &0.72 &0.49 &0.68 &0.49 &0.66 &0.29 &0.66 &0.41\\
        HISEvent &\textbf{0.89} &0.74 &0.88 &0.65 &\textbf{0.90} &\textbf{0.87} &\textbf{0.82} &0.62 &\textbf{0.90} &0.82 &0.78 &0.46 &\textbf{0.88} 
        &\textbf{0.85}\\
        \hline
        DP$_G$-SEMEvent ($\epsilon$=10) &0.33 &0.14 &0.35 &0.32 &0.76 &0.80 &0.59 &0.48 &0.46 &0.38 &0.56 &0.36 &0.51 &0.49 \\
        DP$_G$-SEMEvent ($\epsilon$=15) &0.78 &0.61 &0.60 &0.45 &0.80 &0.80 &0.77 &0.0 &0.64 &0.57 &0.65 &0.39 &0.65 &0.50 \\
        ADP-Spectral Clustering ($\epsilon$=10) &0.01 &0.01 &0.01 &0.01 &0.01 &0.02 &0.05 &0.10 &0.03 &0.09 &0.00 &0.02 &0.04 &0.13 \\
        ADP-Spectral Clustering ($\epsilon$=15) &0.06 &0.02 &0.06 &0.04 &0.02 &0.04 &0.03 &0.04 &0.04 &0.02 &0.14 &0.05 &0.01 &0.02 \\
        \hline
        ADP-SEMEvent ($\epsilon$=10) &0.69 &0.60 &0.57 &0.47 &0.77 &0.80 &0.69 &0.56 &0.56 &0.42 &0.55 &0.35 &0.51 &0.44\\
        ADP-SEMEvent ($\epsilon$=15) &0.73 &0.53 &0.85 &0.76  &0.88 &0.85 &0.81 &0.64 &0.74 &0.74 &0.72 &0.43 &0.71 &0.58\\
        ADP-SEMEvent ($\epsilon$=None) &\textbf{0.89} &\textbf{0.75} &\textbf{0.89} &\textbf{0.79} &\textbf{0.90} &\textbf{0.87} &\textbf{0.82} &\textbf{0.65} &\textbf{0.90} &\textbf{0.88} &\textbf{0.79} &\textbf{0.51} &0.83 &0.71 \\
        \toprule

        \toprule
        \multirow{2}{*}{Methods} &\multicolumn{2}{c|}{$M_{15}$} &\multicolumn{2}{c|}{$M_{16}$} &\multicolumn{2}{c|}{$M_{17}$} &\multicolumn{2}{c|}{$M_{18}$} &\multicolumn{2}{c|}{$M_{19}$} &\multicolumn{2}{c|}{$M_{20}$} &\multicolumn{2}{c}{$M_{21}$}\\
        \cline{2-15}
        &AMI &ARI &AMI &ARI &AMI &ARI &AMI &ARI &AMI &ARI &AMI &ARI &AMI &ARI \\
        \hline
        BERT &0.43 &0.07 &0.71 &0.43 &0.56 &0.22 &0.52 &0.24 &0.59 &0.28 &0.60 &0.32 &0.54 &0.17 \\
        SBERT &0.67 &0.17 &0.78 &0.50 &0.77 &0.35 &0.81 &0.52 &0.83 &0.54 &0.80 &0.52 &0.70 &0.24 \\
        EventX &0.07 &0.01 &0.19 &0.08 &0.18 &0.12 &0.16 &0.08 &0.16 &0.07 &0.18 &0.11 &0.10 &0.10\\
        KPGNN &0.54 &0.17 &0.77 &0.66 &0.68 &0.43 &0.66 &0.47 &0.71 &0.51 &0.68 &0.51 &0.57 &0.20\\
        QSGNN &0.55 &0.17 &0.76 &0.65 &0.69 &0.44 &0.68 &0.48 &0.70 &0.50 &0.69 &0.51 &0.58 &0.21  \\
        HISEvent &0.72 &0.27 &0.87 &0.83 &\textbf{0.81} &0.56 &0.80 &\textbf{0.70} &\textbf{0.87} &0.63 &0.81 &\textbf{0.69} &0.69 &0.45\\
        \hline
        DP$_G$-SEMEvent ($\epsilon$=10) &0.42 &0.32 &0.73 &0.75 &0.55 &0.56 &0.48 &0.40 &0.60 &0.48 &0.51 &0.41 &0.57 &0.33 \\
        DP$_G$-SEMEvent ($\epsilon$=15) &0.56 &0.33 &0.82 &0.78 &0.69 &\textbf{0.58} &0.66 &0.58 &0.86 &0.56 &0.68 &0.58 &0.57 &0.33 \\
        ADP-Spectral Clustering ($\epsilon$=10) &0.10 &0.20 &0.02 &0.03 &0.04 &0.05 &0.03 &0.10 &0.09 &0.01 &0.01 &0.02 &0.03 &0.03\\
        ADP-Spectral Clustering ($\epsilon$=15) &0.02 &0.03 &0.02 &0.03 &0.06 &0.04 &0.04 &0.02 &0.18 &0.10 &0.03 &0.01 &0.05 &0.05  \\
        \hline
        ADP-SEMEvent ($\epsilon$=10) &0.63 &0.30 &0.83 &0.80 &0.56 &0.57 &0.54 &0.43 &0.72 &0.63 &0.66 &0.64 &0.45 &0.40 \\
        ADP-SEMEvent ($\epsilon$=15) &0.60 &\textbf{0.35} &0.84 &\textbf{0.84} &0.74 &0.57 &0.73 &0.65 &0.84 &0.67 &0.76 &0.64 &0.66 &0.42 \\
        ADP-SEMEvent ($\epsilon$=None) &\textbf{0.75} &0.30 &\textbf{0.88} &\textbf{0.84} &0.79 &\textbf{0.58} &\textbf{0.81} &\textbf{0.70} &0.86 &\textbf{0.68} &\textbf{0.82} &\textbf{0.69} &\textbf{0.71} &\textbf{0.46} \\
        \toprule
    \end{tabular}}
\label{r_event2012}
\end{table*}
\section{Experiments}

In this section, we evaluate the event detection and privacy protection capabilities of ADP-SEMEvent.
Specifically, our goal is to answer the following questions:
\textbf{Q1}: How does ADP-SEMEvent perform compared to other methods?
\textbf{Q2}: What is the capability of ADP-SEMEvent in privacy protection?
\textbf{Q3}: What is the impact of privacy budget $\epsilon$ on ADP-SEMEvent's results?
\textbf{Q4}: How is the new 2D SE minimization algorithm executed?
\textbf{Q5}: How does the adaptive differential privacy strategy work?

\begin{table*}[!t]
    \vspace{-2mm}
    \caption{Open-set results on Event2018.}
    \vspace{-2mm}
    \centering
    \renewcommand\arraystretch{0.94}
    \resizebox{\linewidth}{!}{
    \begin{tabular}{c||c c|c c|c c|c c|c c|c c|c c|c c}
        \toprule
        \multirow{2}{*}{Methods} &\multicolumn{2}{c|}{$M_1$} &\multicolumn{2}{c|}{$M_2$} &\multicolumn{2}{c|}{$M_3$} &\multicolumn{2}{c|}{$M_4$} &\multicolumn{2}{c|}{$M_5$} &\multicolumn{2}{c|}{$M_6$} &\multicolumn{2}{c|}{$M_7$} &\multicolumn{2}{c}{$M_8$}\\
        \cline{2-17}
        &AMI &ARI &AMI &ARI &AMI &ARI &AMI &ARI &AMI &ARI &AMI &ARI &AMI &ARI  &AMI &ARI \\
        \hline
        BERT &0.42 &0.16 &0.44 &0.21 &0.22 &0.22 &0.41 &0.17 &0.56 &0.31 &0.49 &0.23 &0.49 &0.23 &0.50 &0.24 \\
        SBERT &0.60 &0.20 &0.61 &0.29 &0.63 &0.34 &0.60 &0.23 &0.76 &0.47 &0.73 &0.41 &0.65 &0.29 &0.75 &0.50 \\
        EventX &0.11 &0.02 &0.12 &0.02 &0.11 &0.01 &0.14 &0.06 &0.24 &0.13 &0.15 &0.08 &0.12 &0.02 &0.21 &0.09\\
        KPGNN &0.54 &0.17 &0.55 &0.18 &0.55 &0.15 &0.55 &0.17 &0.57 &0.21 &0.57 &0.21 &0.61 &0.30 &0.57 &0.20 \\
        QSGNN &0.56 &0.18 &0.57 &0.19 &0.56 &0.17 &0.59 &0.18 &0.59 &0.23 &0.59 &0.21 &0.63 &0.30 &0.55 &0.19 \\
        HISEvent &\textbf{0.78} &0.55 &\textbf{0.78} &0.60 &\textbf{0.75} &0.50 &\textbf{0.73} &0.48 &\textbf{0.79} &0.57 &0.81 &0.56 &\textbf{0.85} &\textbf{0.67} &\textbf{0.89} &0.80\\
        \hline
        DP$_G$-SEMEvent ($\epsilon$=10) &0.40 &0.35 &0.45 &0.43 &0.36 &0.24 &0.40 &0.23 &0.39 &0.40 &0.44 &0.42 &0.41 &0.35 &0.30 &0.26  \\
        DP$_G$-SEMEvent ($\epsilon$=15) &0.52 &0.47 &0.55 &0.53 &0.51 &0.38 &0.52 &0.38 &0.55 &0.47 &0.57 &0.63 &0.60 &0.53 &0.52 &0.44 \\
        ADP-Spectral Clustering ($\epsilon$=10) &0.01 &0.03 &0.02 &0.02 &0.01 &0.02 &0.05 &0.02 &0.10 &0.09 &0.02 &0.04 &0.02 &0.01 &0.08 &0.02\\
        ADP-Spectral Clustering ($\epsilon$=15) &0.02 &0.02 &0.05 &0.03 &0.02 &0.02 &0.06 &0.07 &0.03 &0.04 &0.03 &0.04 &0.09 &0.08 &0.05 &0.04\\
        \hline
        ADP-SEMEvent ($\epsilon$=10) &0.48 &0.41 &0.57 &0.55 &0.45 &0.38 &0.53 &0.26 &0.56 &0.44 &0.56 &0.51 &0.58 &0.41 &0.41 &0.37 \\
        ADP-SEMEvent ($\epsilon$=15) &0.56 &0.49 &0.68 &0.65 &0.61 &0.45 &0.62 &0.43 &0.73 &\textbf{0.71} &0.67 &0.71 &0.66 &0.55 &0.66 &0.59\\
        ADP-SEMEvent ($\epsilon$=None) &0.75 &\textbf{0.60} &0.77 &\textbf{0.70 }&\textbf{0.75} &\textbf{0.54} &\textbf{0.73} &\textbf{0.51} &\textbf{0.79} &0.58 &\textbf{0.82} &\textbf{0.73} &\textbf{0.85} &0.65 &\textbf{0.89} &\textbf{0.82}\\
        \toprule

        \toprule
        \multirow{2}{*}{Methods} &\multicolumn{2}{c|}{$M_9$} &\multicolumn{2}{c|}{$M_{10}$} &\multicolumn{2}{c|}{$M_{11}$} &\multicolumn{2}{c|}{$M_{12}$} &\multicolumn{2}{c|}{$M_{13}$} &\multicolumn{2}{c|}{$M_{14}$} &\multicolumn{2}{c|}{$M_{15}$}  &\multicolumn{2}{c}{$M_{16}$} \\
        \cline{2-17}
        &AMI &ARI &AMI &ARI &AMI &ARI &AMI &ARI &AMI &ARI &AMI &ARI &AMI &ARI  &AMI &ARI \\
        \hline
        BERT &0.42 &0.17 &0.46 &0.19 &0.48 &0.18 &0.54 &0.32 &0.40 &0.18 &0.52 &0.27 &0.53 &0.28 &0.43 &0.21 \\
        SBERT &0.63 &0.23 &0.72 &0.39 &0.70 &0.31 &0.76 &0.54 &0.65 &0.34 &0.68 &0.43 &0.71 &0.40 &0.65 &0.25\\
        EventX &0.16 &0.07 &0.19 &0.07 &0.18 &0.06 &0.20 &0.09 &0.15 &0.06 &0.22 &0.11 &0.22 &0.11 &0.10 &0.01\\
        KPGNN &0.46 &0.10 &0.56 &0.18 &0.53 &0.16 &0.56 &0.17 &0.60 &0.28 &0.65 &0.43 &0.58 &0.25 &0.50 &0.13 \\
        QSGNN &0.46 &0.13 &0.58 &0.19 &0.59 &0.20 &0.59 &0.20 &0.58 &0.27 &0.67 &0.44 &0.61 &0.27 &0.50 &0.13 \\
        HISEvent &0.76 &0.52 &\textbf{0.80} &\textbf{0.52} &0.81 &0.50 &\textbf{0.85} &\textbf{0.73} &\textbf{0.89} &\textbf{0.83} &\textbf{0.89} &0.81 &0.83 &0.70 &\textbf{0.68} &0.30 \\
        \hline
        DP$_G$-SEMEvent ($\epsilon$=10) &0.22 &0.12 &0.28 &0.22 &0.35 &0.19 &0.45 &0.36 &0.44 &0.40 &0.45 &0.37 &0.50 &0.42 &0.26 &0.19 \\
        DP$_G$-SEMEvent ($\epsilon$=15) &0.36 &0.12 &0.50 &0.42 &0.52 &0.35 &0.64 &0.67 &0.56 &0.60 &0.60 &0.51 &0.64 &0.63 &0.50 &0.25 \\
        ADP-Spectral Clustering ($\epsilon$=10) &0.03 &0.02 &0.02 &0.03 &0.03 &0.02 &0.02 &0.02 &0.02 &0.01 &0.07 &0.02 &0.04 &0.01 &0.06 &0.03\\
        ADP-Spectral Clustering ($\epsilon$=15) &0.05 &0.06 &0.02 &0.04 &0.03 &0.03 &0.05 &0.07 &0.01 &0.03 &0.06 &0.02 &0.02 &0.09 &0.04 &0.06 \\
        \hline
        ADP-SEMEvent ($\epsilon$=10) &0.36 &0.22 &0.43 &0.26 &0.38 &0.22 &0.47 &0.37 &0.46 &0.43 &0.47 &0.40 &0.58 &0.58 &0.39 &0.23\\
        ADP-SEMEvent ($\epsilon$=15) &0.53 &0.48 &0.60 &\textbf{0.52} &0.63 &0.43 &0.70 &\textbf{0.73} &0.66 &0.69 &0.71 &0.69 &0.76 &0.68 &0.52 &0.27  \\ 
        ADP-SEMEvent ($\epsilon$=None) &\textbf{0.79} &\textbf{0.53} &0.75 &0.50 &\textbf{0.83} &\textbf{0.55} &\textbf{0.85} &\textbf{0.73} &\textbf{0.89} &\textbf{0.83} &\textbf{0.90} &\textbf{0.82} &\textbf{0.85} &\textbf{0.72} &0.67 &\textbf{0.32}   \\
        \toprule
    \end{tabular}}
\label{r_event2018}
\end{table*}
\begin{table}[!t]
    \vspace{-2mm}
    \caption{Close-set results on Event2012 and Event2018.}
    \vspace{-2mm}
    \centering
    \renewcommand\arraystretch{0.93}
    % \setlength{\tabcolsep}{1.5mm}
    % \resizebox{\linewidth}{!}{
    \begin{tabular}{c||c c|c c}
        \toprule
        \multirow{2}{*}{Methods} &\multicolumn{2}{c|}{Event2012} &\multicolumn{2}{c}{Event2018} \\
        \cline{2-5}
        &AMI &ARI &AMI &ARI \\
        \hline
        BERT &0.43 &0.12 &0.34 &0.05 \\
        SBERT &0.73 &0.17 &0.62 &0.11 \\
        EventX &0.19 &0.05 &0.16 &0.03 \\
        KPGNN &0.52 &0.22 &0.44 &0.15 \\
        QSGNN &0.53 &0.22 &0.44 &0.16 \\
        HISEvent &\textbf{0.81} &0.50 &0.47 &0.31 \\
        \hline
        DP$_G$-SEMEvent ($\epsilon$=10) &0.50 &0.20 &0.20 &0.13  \\
        DP$_G$-SEMEvent ($\epsilon$=15) &0.60 &0.28 &0.22 &0.13  \\
        ADP-Spectral Clustering ($\epsilon$=10) &0.10 &0.05 &0.08 &0.03 \\
        ADP-Spectral Clustering ($\epsilon$=15) &0.09 &0.04 &0.08 &0.02 \\
        \hline
        ADP-SEMEvent ($\epsilon$=10) &0.51 &0.41 &0.20 &0.14 \\
        ADP-SEMEvent ($\epsilon$=15) &0.65 &0.49 &0.25 &0.15 \\
        ADP-SEMEvent ($\epsilon$=None)  &\textbf{0.81} &\textbf{0.52} &\textbf{0.49} &\textbf{0.33} \\
        \toprule
    \end{tabular}
\label{r_close}
\end{table}
\subsection{Experimental Setup}\label{Exp_set}
\textbf{Datasets}.
We experiment on two large public Twitter datasets. 
\textbf{Event2012} ~\cite{mcminn2013building} has 68,841 English messages, including 503 events; \textbf{Event2018} ~\cite{mazoyer2020french} has 64,516 French messages, including 257 events. 
We employ the time-splitting method proposed by Cao et al. ~\cite{cao2024hierarchical} and Ren et al. ~\cite{ren2022known} to partition the dataset. Specifically, Event2012 is divided into $M_0$ , $\dots$ ,  $M_{21}$, a total of 22 blocks; Event2018 is divided into $M_0$ , $\dots$ ,  $M_{16}$, a total of 17 blocks. 
Our model is unsupervised and does not require a training set, but some baselines are supervised, so the first day is the training set.  

\textbf{Baselines}. 
We select 8 baselines. 
\textbf{BERT} ~\cite{devlin2019bert}, and \textbf{SBERT} ~\cite{reimers2019sentence}, which transform messages into message embeddings and then employ K-means clustering for social event detection, requiring the total number of events to be specified in advance; 
\textbf{EventX} ~\cite{liu2020story}, which utilizes community detection methods for unsupervised social event detection; 
\textbf{KPGNN} ~\cite{cao2021knowledge}, a supervised GNN detection method; 
\textbf{QSGNN} ~\cite{ren2022known}, a supervised method based on pseudo labels, which performs well in open environments; 
\textbf{HISEvent} ~\cite{cao2024hierarchical}, which utilizes structural entropy for unsupervised clustering to achieve social event detection, is the current SOTA. 
Additionally, \textbf{DP$_G$-SEMEvent}, a variant of our method, utilizes the Gaussian mechanism ~\cite{liu2018generalized} for noise perturbation without employing a mixed sensitivity strategy; 
\textbf{ADP-Spectral Clustering}, running spectral clustering ~\cite{von2007tutorial} directly on our published privacy graph. 
Our results are the average of 5 runs.

\textbf{Hyperparameters}. 
For ADP-SEMEvent, we set the $\epsilon$ to three values: 10, 15, and None (indicating no privacy protection strategy). 
In open environments, the parameter $q$ is set to 400, while in closed environments, it is set to 300, and the same applies to the HISEvent, DP$_G$-SEMEvent, and DP$_G$-SEMEvent. 
For KPGNN, QSGNN, and EventX, we adopt the settings reported in the original papers. 

\textbf{Evaluation Metrics}. 
We measure adjusted mutual information (AMI) and adjusted rand index (ARI). Both metrics range from -1 to 1, where a value closer to 1 indicates better agreement between the predicted and true clustering.

\begin{table*}[!t]
    \vspace{-2mm}
    \caption{The results of attribute inference attack experiment.}
    \vspace{-2mm}
    \centering
    \renewcommand\arraystretch{0.95}
    \resizebox{\linewidth}{!}{
    \begin{tabular}{c||c c|c c||c c|c c||c c|c c}
        \toprule
        \multirow{2}{*}{Methods} &\multicolumn{2}{c|}{PERSON (SVM)} &\multicolumn{2}{c||}{GPE (SVM)} &\multicolumn{2}{c|}{PERSON (LR)} &\multicolumn{2}{c||}{GPE (LR)} &\multicolumn{2}{c|}{PERSON (MLP)} &\multicolumn{2}{c}{GPE (MLP)} \\
        \cline{2-13}
        &F1 &Precision &Accuracy &AUC &F1 &Precision &Accuracy &AUC &F1 &Precision &Accuracy &AUC \\
        \hline
        BERT &0.72 &0.68 &0.86 &0.98 &0.75 &0.68 &0.87 &0.98 &0.74 &0.76 &0.89 &0.98 \\
        SBERT &0.71 &0.78 &0.76 &0.96 &0.68 &0.70 &0.83 &0.97 &0.73 &0.69 &0.81 &0.97 \\
        HISEvent &0.42 &0.40 &0.34 &0.36 &0.41 &0.40 &0.32 &0.58 &0.29 &0.56 &0.31 &0.56 \\
        \hline
        DP$_G$-SEMEvent ($\epsilon$=10) &\textbf{0.33} &\textbf{0.34} &0.33 &0.33 &\textbf{0.32} &\textbf{0.35} &\textbf{0.31} &0.50 &0.25 &\textbf{0.45} &\textbf{0.21} &0.46 \\
        DP$_G$-SEMEvent ($\epsilon$=15) &0.39 &0.39 &0.30 &\textbf{0.31} &0.38 &0.40 &0.33 &0.52 &\textbf{0.21} &0.51 &0.22 &0.51\\
        \hline
        ADP-SEMEvent ($\epsilon$=10) &\textbf{0.33} &0.35 &0.32 &0.34 &0.35 &\textbf{0.35} &\textbf{0.31} &0.49 &\textbf{0.21} &0.46 &\textbf{0.21} &\textbf{0.45} \\
        ADP-SEMEvent ($\epsilon$=15) &0.40 &0.38 &\textbf{0.29} &0.32 &0.38 &0.40 &0.32 &0.52 &0.22 &0.51 &0.22 &0.51\\
        \toprule
    \end{tabular}}
\label{t_Security}
\end{table*}
\subsection{Overall Performance (Q1)}\label{Open}
Tables ~\ref{r_event2012} and ~\ref{r_event2018} present the evaluation results of Event2012 and Event2018 in an open environment, demonstrating the superior performance of ADP-SEMEvent. 
For instance, the results from Event2012 show that in most cases, when $\epsilon = \text{None}$, ADP-SEMEvent outperforms the current SOTA model HISEvent. 
Optimally, the ARI in $M_9$ increases from 0.65 to 0.79, representing a 21.5$\%$ improvement. 
When $\epsilon = 10$ or $15$, compared to the privacy-preserving models DP$_G$-SEMEvent and ADP-Spectral Clustering, ADP-SEMEvent surpasses the baselines on all metrics. 
The comparison with DP$_G$-SEMEvent indicates that our adaptive differential privacy strategy effectively controls noise, and the comparison with ADP-Spectral Clustering highlights that our Algorithm ~\ref{Algorithm 2} can cluster private graphs more effectively, making it a more efficient unsupervised method. 
In summary, ADP-SEMEvent has three characteristics: 
1) it does not require the specification of the total number of events, making it unsupervised; 
2) it emphasizes privacy protection; 
3) it exhibits excellent performance and robustness. 
No baseline simultaneously satisfies all of these criteria. 

Table~\ref{r_close} shows the evaluation results for Event2012 and Event2018 in a closed environment. 
When $\epsilon$=None on large datasets, our ADP-SEMEvent achieves SOTA performance. 
Performance decreases for $\epsilon$=10 and $\epsilon$=15, but it still outperforms DP$_G$-SEMEvent and ADP-Spectral Clustering. 
Test results for Event2012 indicate that ADP-SEMEvent remains competitive; however, the results for Event2018 show lower performance, possibly due to the closer proximity of local sensitivity and global sensitivity in the Event2018 dataset.

\subsection{Security Analysis (Q2)}\label{Security}
We conduct a security analysis for all models. 
For Eventx, KPGNN, and QSGNN, are supervised. 
Generally, machine learning models tend to exhibit memorization, enabling malicious attackers to leverage membership inference attacks to obtain data from the training set ~\cite{salem2018ml}.
For ADP-SEMEvent and other baselines, unsupervised clustering methods are used, eliminating the need for training data. 
Consequently, the risk of membership inference attacks is greatly reduced. 
However, attackers can still design attribute inference attacks ~\cite{gong2018attribute,zhao2021feasibility} to deduce sensitive information about messages, such as relationships and locations. 

To assess ADP-SEMEvent's protection capabilities, we design attribute inference attack experiments, selecting *Person* and *GPE* (Geopolitical Entity) as the attack target. 
We manually select 535 pieces of data containing sensitive attributes from Event2012. 
To comprehensively assess security, we design binary and multi-class attack scenarios. 
Specifically, for *Person*, we only consider whether the data contains the *Person* attribute, making it a binary classification problem. 
We use F1 and precision as evaluation metrics. 
For *GPE*, which includes 8 categories, it is a multi-classification problem. 
We use accuracy and AUC as evaluation metrics. 
For BERT and SBERT, we directly use embeddings as features of the messages; for other models, we use a simple Node2Vec ~\cite{grover2016node2vec} model to obtain node embeddings from the graph; 30\% of the data is used as the training set.
We choose support vector product (SVM), logistic regression (LR), and multi-layer perceptron (MLP) as attack models

Our test results are shown in Table ~\ref{t_Security}. 
Since BERT and SBERT directly convert messages into embeddings, sensitive information is directly exposed in the embeddings, leading to the worst evaluation metrics. 
Although HISEvent's probability of a successful attack is lower than that of directly using embeddings, its privacy protection capabilities remain limited due to a lack of further processing of sensitive information. 
For ADP-SEMEvent and DP$_G$-SEMEvent, the attack model's effectiveness on these models is relatively low, indicating that the noise provided some degree of protection. 
However, ADP-SEMEvent outperforms DP$_G$-SEMEvent in event detection tasks, demonstrating its higher application value. 
% Moreover, because our model adheres to differential privacy, even if the attacker infers some knowledge, they must consider that it was inferred under a noisy background, making it impossible to determine its authenticity.

\begin{figure}[!t]
    \centering
    \includegraphics[width=0.95\linewidth]{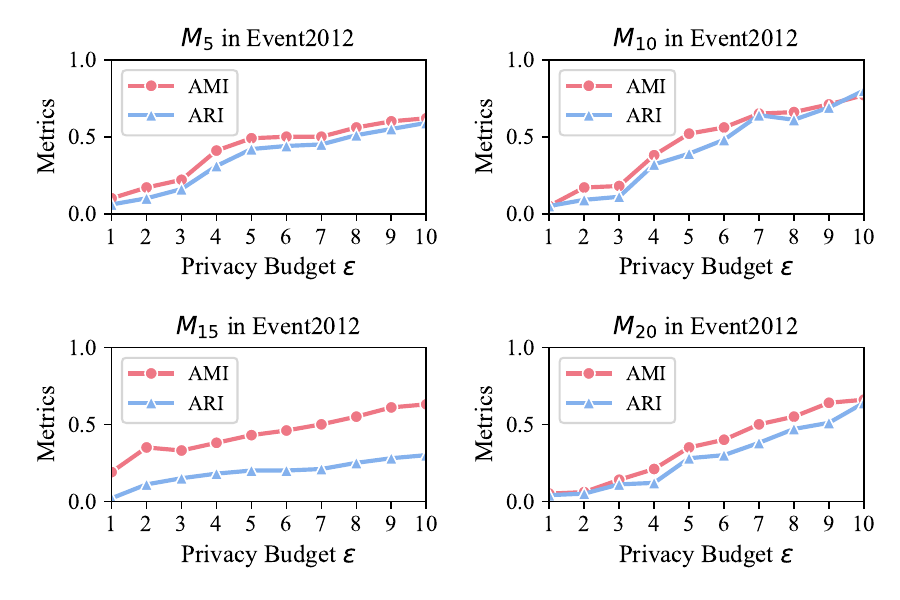}
    \vspace{-2mm}
    \caption{Sensitivity analysis results of the model to privacy budget $\epsilon$.}
    \label{Hyperparameter_Sensitivity}
    \vspace{-2mm}
\end{figure}
\subsection{Hyperparameter Sensitivity (Q3)}\label{Hyperparameter}
To validate the sensitivity of ADP-SEMEvent to the privacy budget $\epsilon$ parameter, we conduct tests using Event2012 in an open environment. 
We incrementally increase the value of the privacy budget $\epsilon$ from 1 to 10 with a step size of 1, then run the ADP-SEMEvent to observe its final performance. 
This test is essential as it provides a straightforward reflection of the relationship between the privacy budget $\epsilon$ and the model performance, aiding algorithm users in balancing privacy protection levels with detection performance in practical applications.
Our test results are illustrated in Figure ~\ref{Hyperparameter_Sensitivity}. 
Analysis of the 4 cases indicates that the privacy budget $\epsilon$ effectively controls the noise level. 
A smaller $\epsilon$ adds more noise, resulting in lower data utility, while a larger $\epsilon$ results in less noise and higher model performance.

\begin{figure}[!t]
    \centering
    \includegraphics[width=0.99\linewidth]{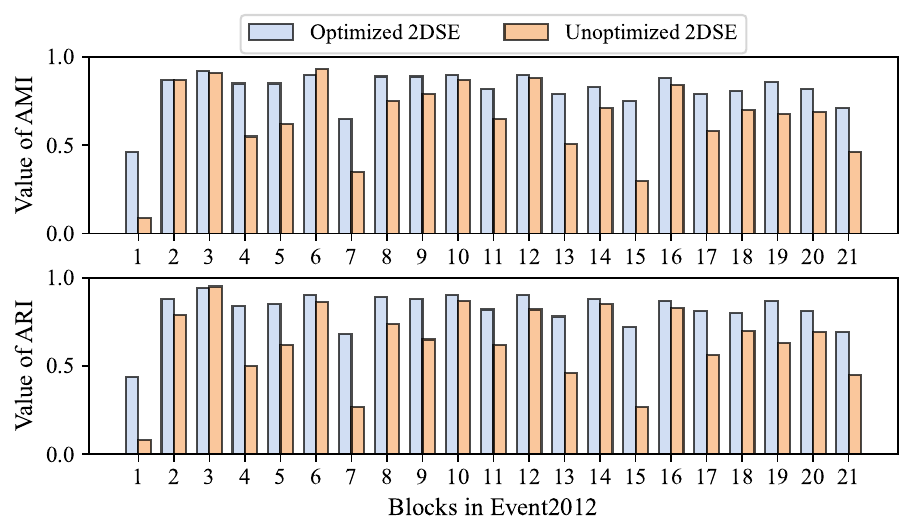}
    \vspace{-2mm}
    \caption{Validity verification results of Optimized Hierarchical 2D SE Minimization. ($\epsilon$ = None) }
    \label{Ablation2_img}
    \vspace{-2mm}
\end{figure}

\subsection{Ablation Study (Q4)}\label{Ablation}
We compare the model's performance before and after optimization to validate the effectiveness of the 2D SE minimization algorithm based on optimal subgraphs. 
As illustrated in Figure ~\ref{Ablation2_img}, in most cases, the algorithm's performance after optimization surpasses that of the unoptimized method. 
This demonstrates that a well-initialized partition aids the subsequent iterative process.

\subsection{Noise Analysis (Q5)}\label{Noise}
\begin{figure}[!tb]
    \centering
    \includegraphics[width=0.95\linewidth]{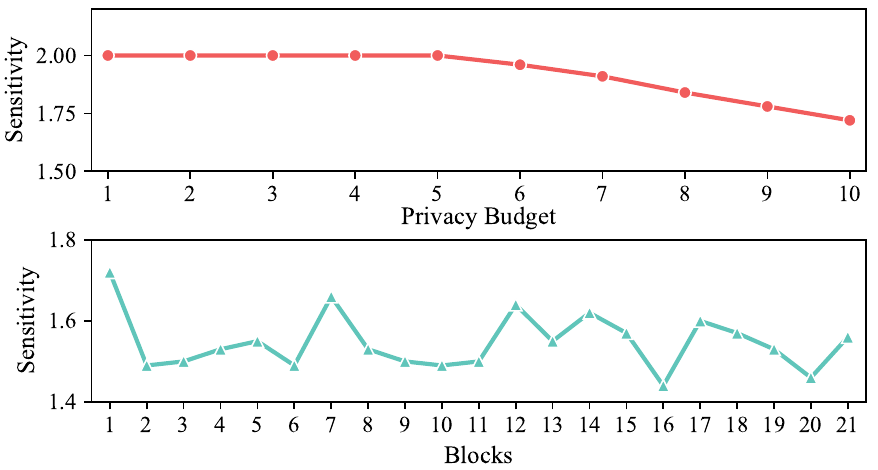}
    \vspace{-3mm}
    \caption{The results of the noise analysis. Above: The impact of privacy budget on noise (Tested on M16, Event2012). Below: The impact of blocks on noise (Privacy budget is set to 15).}
    \label{Noise Analysis}
    \vspace{-2mm}
\end{figure}

To verify the noise level control by the adaptive sensitivity strategy, we use sensitivity as an indicator of noise level, where a higher sensitivity indicates greater perturbation to the original data. 
We conduct statistical analyses of noise levels under different privacy budgets and across different datasets. 
The results are shown in Figure ~\ref{Noise Analysis}.
We find that when the privacy budget $\epsilon \geq 5$, our mixed sensitivity begins to take effect, avoiding noise waste. 
In a continuous 21-day simulation, the mixed sensitivity can determine the most appropriate sensitivity based on the current day's situation, balancing performance and privacy.

\section{Related Work}

\subsection{Differential Privacy}\label{work_DP}
Differential privacy was first introduced by Dwork et al.~\cite{dwork2006differential}. 
The most commonly used method of differentially private is the Laplace mechanism ~\cite{dwork2006calibrating}, which adds appropriate noise to meet $\epsilon$-DP by analyzing the global sensitivity of the query function. 
Subsequently, Dwork et al. ~\cite{dwork2006our} proposed approximate differential privacy, which relaxes the strict requirement of $\epsilon$-DP by using the Gaussian mechanism to add noise, improving the applicability. 
To address the challenge of computing global sensitivity, Dwork et al. ~\cite{dwork2009Differential} proposed the Propose-Test-Release framework, while Nissim et al. ~\cite{nissim2007smooth} introduced the smooth sensitivity and Sample and Aggregate frameworks. 
Nowadays, due to the growing problem of privacy data leakage, many scholars utilize differential privacy to protect data.
For example, in private clustering ~\cite{imola2023differentially,huang2018optimal}, private PageRank~\cite{epasto2022differentially}, private data release or synthesis ~\cite{yuan2023privgraph,li2023locally,ren2022cross}, and application in recommender system~\cite{wei2022heterogeneous}, among others. 
However, existing DP methods do not effectively adapt to the addition of noise based on each day's events.

\subsection{Social Event Detection}\label{work_event}
Existing research ~\cite{cao2024hierarchical,peng2019fine,peng2022reinforced,ren2022cross} typically formalizes the task of social event detection as clustering relevant messages from social media sequences to represent events. 
% Based on the model input, social event detection methods can generally be divided into the following three categories: 
We divide social event detection methods into three categories based on the model input, i.e., attribute-based, content-based, and Hybrid methods. 
1) Attribute-based methods ~\cite{hu2022event,singh2021burst} require predefined rules or patterns and then identify events by matching rules. 
Their drawback is the inability to capture complex event features accurately, and updating rules for newly emerged event types may be required, resulting in lower performance and robustness. 
2) Content-based methods ~\cite{wang2017neural, Sihem2020Event} mostly rely on natural language processing to model and analyze content. 
Their detection performance heavily depends on content modeling. 
However, due to the complexity and variability of real-world content, methods relying solely on content modeling have not shown good robustness. 
3) Hybrid methods ~\cite{peng2019fine,cao2021knowledge,ren2022known,peng2022reinforced,ren2021transferring,li2023type,li2023event} utilize GNNs for modeling, effectively combining content and attributes into heterogeneous graphs to achieve better performance. 
% However, since GNN contains rich node information, it is easy to be attacked.
However, since GNNs contain rich node information, they may inevitably include sensitive data that attackers should not be able to access, making privacy protection crucial for the models.

\section{Conclusion}

In this work, we recognize the importance of privacy protection and have explored how to conduct social event detection under the requirements of privacy protection. 
We propose a novel unsupervised social event detection framework, ADP-SEMEvent, implementing adaptive privacy protection. 
% Specifically, our method based on optimal subgraphs' structural entropy retains the advantages of graph-based methods in better node interactions. 
The adaptive differentially private strategy we proposed maximizes the utilization of the privacy budget and achieves a good balance between privacy and accuracy. 
At the same time, the unsupervised approach eliminates the need to determine the number of events in advance, providing better robustness in real-world applications. 
Experiments show that ADP-SEMEvent can protect privacy while achieving satisfactory results.
Since PLMs have a significant impact on both performance and privacy, we will explore efficient and private PLMs in future work.

\begin{acks}
This research is supported by the National Key R\&D Program of China through grant 2023YFC3303800, NSFC through grant 62322202, Shijiazhuang Science and Technology Plan Project through grant 231130459A, and Guangdong Basic and Applied Basic Research Foundation through grant 2023B1515120020.
\end{acks}

\newpage

\bibliographystyle{ACM-Reference-Format}
\bibliography{sample-base}

\end{document}